\documentclass[letter,traditabstract,longauth]{aa} % for the letters 
\usepackage{graphicx}
\usepackage{txfonts}
\usepackage{natbib}
\bibpunct{(}{)}{;}{a}{}{,}

\def\ltsima{$\; \buildrel < \over \sim \;$}
\def\simlt{\lower.5ex\hbox{\ltsima}}
\def\gtsima{$\; \buildrel > \over \sim \;$}
\def\simgt{\lower.5ex\hbox{\gtsima}}

\def\cm2{\mbox{$\mbox{cm}^{-2}$}}
\def\cm3{\mbox{$\mbox{cm}^{-3}$}}
\def\h2{\mbox{$_{\mbox{\tiny H2}}$}}
\newcommand{\um}{\hbox{${\mu}$m}}
\begin{document}

\title
{From filamentary clouds to prestellar cores to the stellar IMF:\\ 
Initial highlights from the \emph{Herschel}\thanks{\emph{Herschel} is an ESA space observatory with science
instruments provided by European-led Principal Investigator
consortia and with important participation from NASA.} Gould belt survey\thanks{Appendix A and Figs. 3, 4 are only available in electronic form via \,\,\,\,\,\,\,\,\,\,\,\,\,\,\,\,\,\,\,\,\,\,\,\,\,\,\,
http://www.edpsciences.org}
}

%||||||||||||||||||||||||||||||||||||||||||||||||||||||||||||||||||||||||||||||||||||||||||||||||||||||||||||||||||||||||||||||||||

\author
{ Ph.~Andr\'e\inst{1}
\and A.~Men'shchikov\inst{1}
\and S.~Bontemps\inst{1}
\and V.~K\"onyves\inst{1}
\and F.~Motte\inst{1}
\and N.~Schneider\inst{1}
\and P.~Didelon\inst{1}
\and V.~Minier\inst{1}
\and P.~Saraceno\inst{5}
\and D.~Ward-Thompson\inst{3}
\and J.~Di~Francesco\inst{10}
\and G.~White\inst{18,22}
\and S.~Molinari\inst{5}
\and L.~Testi\inst{17}
\and A.~Abergel\inst{2}
\and M.~Griffin\inst{3}
\and Th.~Henning\inst{11}
\and P.~Royer\inst{7}
\and B.~Mer\'in\inst{13}
\and R.~Vavrek\inst{13}
\and M.~Attard\inst{1}
\and D.~Arzoumanian\inst{1}
\and C. D.~Wilson\inst{19}
\and P.~Ade\inst{3}
\and H.~Aussel\inst{1}
\and J.-P.~Baluteau\inst{4}
\and M.~Benedettini\inst{5}
\and J.-Ph.~Bernard\inst{6}
\and J.A.D.L.~Blommaert\inst{7}
\and L.~Cambr\'esy\inst{8}
\and P.~Cox\inst{9}
\and A.~Di~Giorgio\inst{5}
\and P.~Hargrave\inst{3}
\and M.~Hennemann\inst{1}
\and M.~Huang\inst{12}
\and J.~Kirk\inst{3}
\and O.~Krause\inst{11}
\and R.~Launhardt\inst{11}
\and S.~Leeks\inst{18}
\and J.~Le Pennec\inst{1}
\and J.Z.~Li\inst{12}
\and P. G.~Martin\inst{14}
\and A.~Maury\inst{1}
\and G.~Olofsson\inst{15}
\and A.~Omont\inst{16}
\and N.~Peretto\inst{1}
\and S.~Pezzuto\inst{5}
\and T.~Prusti\inst{21}
\and H.~Roussel\inst{16}
\and D.~Russeil\inst{4}
\and M.~Sauvage\inst{1}
\and B.~Sibthorpe\inst{20}
\and A.~Sicilia-Aguilar\inst{11}
\and L.~Spinoglio\inst{5}
\and C.~Waelkens\inst{7}
\and A.~Woodcraft\inst{20}
\and A.~Zavagno\inst{4}
}

\offprints{pandre@cea.fr}

%||||||||||||||||||||||||||||||||||||||||||||||||||||||||||||||||||||||||||||||||||||||||||||||||||||||||||||||||||||||||||||||||||

\institute
{
Laboratoire AIM, CEA/DSM--CNRS--Universit\'e Paris Diderot, IRFU/Service d'Astrophysique, CEA Saclay, 
91191 Gif-sur-Yvette, France
\and Institut d'Astrophysique Spatiale, CNRS/Universit{\'e} Paris-Sud 11, 91405 Orsay, France 
\and School of Physics and Astronomy, Cardiff University, Queens Buildings, The Parade, Cardiff CF243AA, UK 
\and Laboratoire dÕAstrophysique de Marseille , CNRS/INSU - Universit\'e de Provence, 13388 Marseille cedex 13, France 
\and INAF-Istituto Fisica Spazio Interplanetario, Via Fosso del Cavaliere 100, I-00133 Roma, Italy
\and CESR, Observatoire Midi-Pyr\'en\'ees (CNRS-UPS), Universit\'e de Toulouse, BP 44346, 31028 Toulouse Cedex 04, France
\and Instituut voor Sterrenkunde, K.U.Leuven, Celestijnenlaan 200D, B-3001 Leuven, Belgium
\and CDS, Observatoire de Strasbourg, 11, rue de l'Universit{\'e}, 67000 Strasbourg, France
\and IRAM, 300 rue de la Piscine, Domaine Universitaire, 38406 Saint Martin d'H{\`e}res, France
\and National Research Council of Canada, Herzberg Institute of Astrophysics, Victoria, BC, V9E 2E7, Canada
\and Max-Planck-Institut f\"ur Astronomie, K\"onigstuhl 17, D-69117 Heidelberg, Germany
\and National Astronomical Observatories, Chinese Academy of Sciences, Beijing 100012, China
\and Herschel Science Centre, ESAC, ESA, PO Box 78, Villanueva de la 
Ca\~nada, 28691 Madrid, Spain
\and Canadian Institute for Theoretical Astrophysics, University of Toronto, 60 St. George Street, Toronto, ONM5S3H8, Canada
\and Stockholm Observatory, AlbaNova University Center, Roslagstullsbacken 21, SE-106 91 Stockholm, Sweden
\and Institut d'Astrophysique de Paris, Universit\'e Pierre \&
Marie Curie, 98 bis Boulevard Arago, 75014 Paris, France
\and INAF, Osservatorio Astrofisico di Arcetri, Firenze, Italy
\and Space Science and Technology Department, Rutherford Appleton 
Laboratory, Chilton, Didcot, Oxon OX11 0QX, UK
\and Dept. of Physics \& Astronomy, McMaster University, Hamilton,  
Ontario, L8S 4M1, Canada
\and UK Astronomy Technology Centre, Royal Observatory Edinburgh, Blackford
Hill, EH9 3HJ, UK
\and ESA/ESTEC, P.O. Box 299, 2200 AG Noordwijk, The Netherlands 
\and Department of Physics \& Astronomy, The Open University, Milton Keynes MK7 6AA, UK
}

\date{Received 1 April 2010; accepted 4 May 2010}

%||||||||||||||||||||||||||||||||||||||||||||||||||||||||||||||||||||||||||||||||||||||||||||||||||||||||||||||||||||||||||||||||||

\abstract{We summarize the first results from the 
Gould belt survey,  obtained toward the Aquila Rift and Polaris Flare regions
during the 'science demonstration phase'  of \emph{Herschel}. 
Our 70--500~$\mu$m images  
taken in parallel mode with the SPIRE and PACS
cameras reveal a wealth of filamentary structure, as well as numerous dense cores embedded in the 
filaments. Between $\sim $ 350 and 500 prestellar cores and $\sim $~45--60 Class~0 protostars can be identified in 
the Aquila field, while $\sim 300$ unbound starless cores and no protostars are observed in the Polaris field.
The prestellar core mass function (CMF) derived for the Aquila region bears a strong resemblance to the 
stellar initial mass function (IMF), already confirming the close connection between the CMF and the IMF 
with much better statistics than earlier studies. 
Comparing and contrasting our $Herschel$ results in Aquila and Polaris, we propose 
an observationally-driven scenario for core formation according to which complex networks of long, thin filaments 
form first within molecular clouds, and then the densest 
filaments fragment into a number of prestellar cores via gravitational instability. 
}


\titlerunning{The $Herschel$ Gould belt survey}

\authorrunning{Andr\'e et al.}

\maketitle

%||||||||||||||||||||||||||||||||||||||||||||||||||||||||||||||||||||||||||||||||||||||||||||||||||||||||||||||||||||||||||||||||||

\section{Introduction}
\label{intro}

The $Herschel$ Space Observatory (Pilbratt et al. 2010) offers a unique opportunity to improve our global understanding 
of  the earliest phases of star formation. 
Here, we present first highlights from the Gould belt survey, 
one of the largest key projects with \emph{Herschel} (cf. Andr\'e \& Saraceno 2005), based on extensive far-infrared and 
submillimeter mapping of nearby molecular clouds with both the SPIRE (Griffin et al. 2010) and PACS (Poglitsch et al. 2010) 
bolometer cameras. 
This SPIRE/PACS imaging survey will cover the bulk of the nearest ($d \leq 0.5$~kpc) cloud complexes in the Galaxy, 
which are mostly  located in  the Gould belt, a giant ($\sim 700\, {\rm pc} \times 1000\,{\rm pc}$), flat structure inclined 
by $\sim 20^\circ $ to the Galactic plane (e.g., Guillout 2001).
Since the $\sim 15\arcsec $ angular resolution of $Herschel$ around 
$\lambda \sim 200\, \mu$m is adequate for probing individual ($\sim$~0.01--0.1~pc) star-forming cores up to 
$\sim 0.5$~kpc away, the cloud complexes of the Gould belt  correspond to the volume of Galactic space where  
$Herschel$ imaging can be best used to characterize in detail the earliest stages of star formation. 

The immediate observational objective of the Gould belt survey is to obtain complete samples of 
nearby prestellar cores and Class 0 protostars with well characterized luminosities, 
temperatures, and density profiles, as well as robust core mass functions
and protostar luminosity functions, in a variety of star-forming environments.
An order of magnitude more cold prestellar cores 
than already identified from the ground  are expected to be found in the
entire survey, which should allow us to derive an accurate prestellar core mass
function (CMF) from the pre-brown-dwarf to the intermediate-mass range. 
Thanks to its high sensitivity and large spatial dynamic range, this $Herschel$ survey 
can also probe, for the first time, the link between low-density cirrus-like structures 
in the interstellar medium (ISM) and compact self-gravitating cores. 
The main scientific goal is to elucidate the 
physical mechanisms responsible for the formation of prestellar cores out of the 
diffuse ISM, which is crucial for understanding the origin of the stellar initial 
mass function (IMF).

Our first results, obtained toward the Aquila Rift and Polaris clouds,  
are very promising 
(e.g., K\"onyves et al. 2010, Bontemps et al. 2010, Men'shchikov et al. 2010, Ward-Thompson et al. 2010, and 
Miville-Desch\^enes et al. 2010). 
As discussed in Sect.~4 of this paper, they suggest that prestellar cores result from the 
{\it gravitational fragmentation of filaments} in the cold ISM.

%||||||||||||||||||||||||||||||||||||||||||||||||||||||||||||||||||||||||||||||||||||||||||||||||||||||||||||||||||||||||||||||||||

\section{\emph{Herschel} observations} 
\label{obs}

The $Herschel$ survey was designed to cover the densest portions of the Gould belt with
SPIRE at 250--500~$\mu$m and PACS at 100--160~$\mu$m.
The observational goal is to make a complete, homogeneous mapping of the $A_V > 3 $
regions with SPIRE and  the $A_V > 6 $  regions with PACS, and representative areas at $A_V \sim $~1--3
levels with both instruments. 
The survey sensitivity (better than $A_V \sim 1$ at the 5$\sigma$ level) will allow us to probe the 
structure of nearby molecular clouds {\it down to the level of the 
interface with their atomic gas envelopes}.
The 15 target clouds span a range of physical conditions, from active, cluster-forming complexes to
quiescent regions with lower or no star formation activity\footnote{See http://gouldbelt-herschel.cea.fr/ for the list of target fields.}.

Our $Herschel$ mapping consists of  two steps:

(1) A wide-field SPIRE/PACS survey of a total surface area $\sim 160\, \rm{deg}^2$ 
using the so-called ``parallel mode'' with a scanning speed of 60\arcsec/s and the PACS 70~$\mu$m and 160~$\mu$m bands. 
In this first step, the main goal is to acquire adequate SPIRE 250--500~$\mu$m data for all of the 
target regions. 
The PACS data acquired simultaneously  
yield very useful information at 70~$\mu$m and 160~$\mu$m 
through most of the SPIRE survey, but do not have optimal (diffraction-limited) angular resolution. 
We selected the 70~$\mu$m filter for the blue band of PACS to obtain a good diagnostic 
of the presence of embedded protostars (cf. Dunham et al. 2008)  throughout the mapped regions.

(2) A dedicated PACS-only survey of a total surface area $\sim 65\, \rm{deg}^2$, 
observing  the 100~$\mu$m and 160~$\mu$m bands with a scanning speed of 20\arcsec/s. 
This second step is supposed to provide data of optimal sensitivity and resolution 
at 100~$\mu$m and 160~$\mu$m.

The Aquila Rift  and Polaris flare regions were observed in parallel mode with both SPIRE and PACS (step 1) 
during the science demonstration phase  of $Herschel$.  
The corresponding observations are described in detail by K\"onyves et al. (2010), Bontemps et al. (2010), Men'shchikov et al. (2010), and 
Ward-Thompson et al. (2010) 
in this special A\&A issue.  
The Polaris flare field is a high-latitude {\it translucent} cloud with little to no star formation at $d \sim 150$~pc (e.g., Heithausen et al. 2002), and 
is expected to have the {\it lowest level of background cloud emission and cirrus confusion noise} in the entire Gould belt survey.
At the other extreme, the Aquila field is a very active star-forming complex at $d \sim 260$~pc 
%(e.g. Strai{\v z}ys et al. 2003, 
(e.g., Gutermuth et al. 2008), and is expected to have the {\it highest level of background cloud emission and cirrus confusion noise} 
in the whole survey. 
Sensitive Galactic far-IR imaging surveys such as the one discussed here are not limited by instrumental sensitivity but 
by confusion arising from small-scale cirrus/cloud structure (cf. Roy et al. 2010 and references therein).
%(e.g. Gautier et al. 1992, Kiss et al. 2001). 
The levels of background cloud fluctuations observed in the Aquila and Polaris fields allow us to roughly estimate the whole 
range of cirrus confusion noise levels that will affect the Gould belt survey.
In the Aquila field, the observed rms level of cirrus or ``structure'' noise (as measured on the typical scales of dense cores) 
ranges from $\sim$~30~mJy/18\arcsec-beam  to $\sim$~300~mJy/18\arcsec-beam at $\lambda = 250\, \mu$m, which is 
a factor of $\sim $~3--30 above the rms instrumental noise.
In contrast, in the Polaris field, the rms level of emission fluctuations measured in the SPIRE $250\, \mu$m map 
is only $\sim$~10--30~mJy/18\arcsec-beam, which is very close to the rms sensitivity level expected after two cross-scans 
in parallel mode at 60\arcsec /s.
Assuming the dust opacity of Hildebrand (1983) and the median dust temperatures derived from our $Herschel$ maps, 
the above surface-brightness sensitivity levels translate into $5\sigma$ column-density detection thresholds of 
$N\h2 \la  3 \times 10^{20}\, {\rm cm}^{-2} $ in Polaris and $N\h2 \la 10^{21}\, {\rm cm}^{-2} $ in Aquila 
(in agreement with the column-density maps shown by 
%Men'shchikov et al. 2010 
Ward-Thompson et al. 2010 and K\"onyves et al. 2010).
This high column-density sensitivity, coupled with the unprecedented surface brightness and spatial 
dynamic ranges  of our $Herschel$ mapping, allows us to probe, for the first time, the physical connection 
between the structure of the diffuse ISM and the formation of prestellar cores (cf. Sect.~4).
% below).

The corresponding mass sensitivities for typical prestellar/protostellar cores are $\sim 0.3\, M_\odot $ 
(85\% completeness level) in Aquila ($d = 260$~pc) 
%, but see Appendix~A) 
and $\sim 0.01\, M_\odot $  (85\% completeness level) in Polaris ($d = 150$~pc). 
These completeness numbers were estimated by performing Monte-Carlo simulations as described by K\"onyves et al. (2010)
in the case of Aquila. The initial results obtained in Aquila and Polaris therefore confirm that the 
completeness level of our $Herschel$ census for prestellar cores will 
%be well into 
reach the pre-brown dwarf mass regime 
($M_{\rm core} < 0.08\, M_\odot $) in the nearest molecular clouds ($d \sim 150$~pc) of the Gould belt.

\begin{figure*}
\centering
%\centerline{\hspace*{-0.1cm}\resizebox{0.46 \hsize}{!}{\includegraphics[angle=270]{14666fig1a.eps}}
%\hspace*{0.8cm}\resizebox{0.44\hsize}{!}{\includegraphics[angle=270]{14666fig1b.eps}}}
\centerline{\hspace*{-0.1cm}\resizebox{0.46 \hsize}{!}{\includegraphics[angle=0]{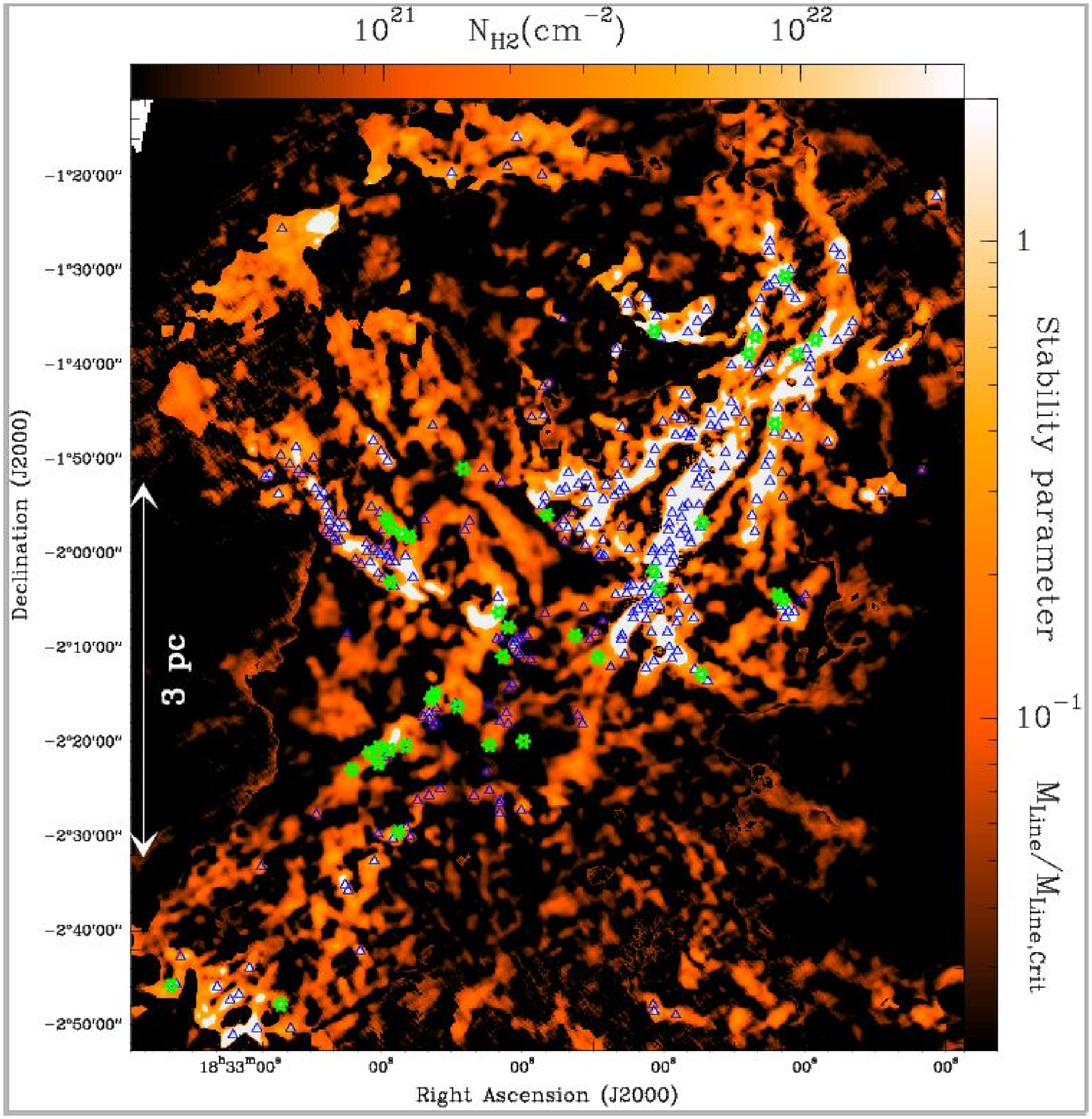}}
\hspace*{0.8cm}\resizebox{0.44\hsize}{!}{\includegraphics[angle=0]{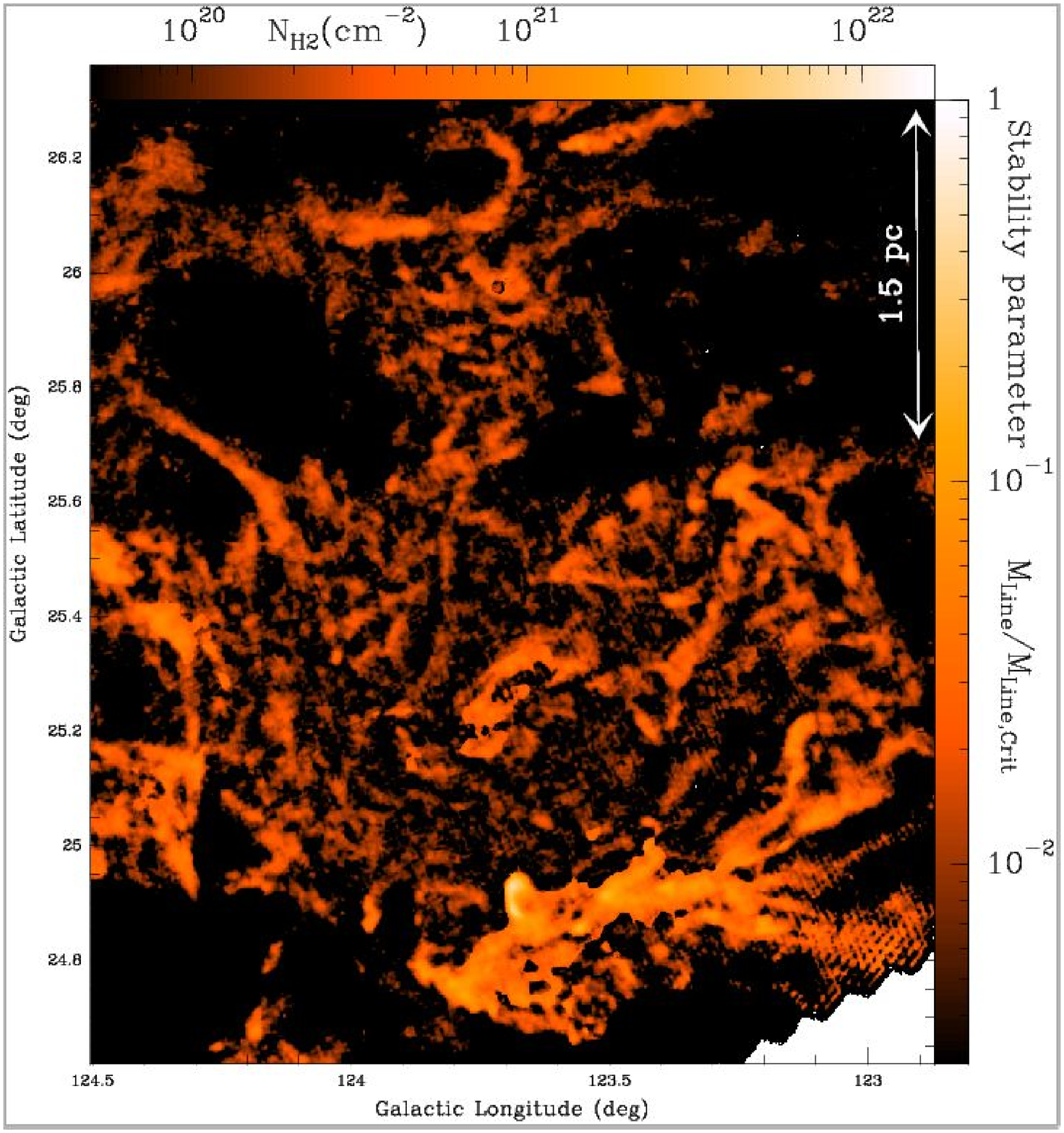}}}
%\centerline{\hspace*{-0.1cm}\resizebox{0.46 \hsize}{!}{\includegraphics[angle=270]{../spire/sf_sag/sd_phase/spire_aquila/aquila_main_filaments.eps}}
%\hspace*{0.8cm}\resizebox{0.44\hsize}{!}{\includegraphics[angle=270]{../spire/sf_sag/sd_phase/spire_polaris/polaris_filaments_coldens.eps}}}
\caption{Column density maps of two subfields in Aquila (left) and Polaris (right) derived from our SPIRE/PACS data.
The contrast of the filaments with respect to the non-filamentary background has been enhanced using a curvelet transform 
as described in 
%Men'shchikov et al. (2010) and 
Appendix~A.
Given the typical width $\sim $~10000~AU of the filaments, these column density maps are equivalent 
to {\it maps of the mass per unit length along the filaments}. 
The color scale shown on the right of each panel is given in approximate units of the critical line mass of Inutsuka \& Miyama (1997) 
as discussed in Sect.~4. 
The areas where the filaments have a mass per unit length larger than half the critical value and are thus likely gravitationally unstable 
have been highlighted in white. 
The maximum line mass observed in the Polaris region is only $\sim 0.45\,  \times$ the critical value, suggesting that the Polaris filaments are stable 
and unable to form stars at the present time. 
The candidate Class~0 protostars and bound prestellar cores identified in 
Aquila by Bontemps et al. (2010) and K\"onyves et al. (2010) 
are shown as green stars and blue triangles, respectively. Note the good correspondence between 
the spatial distribution of the bound cores/protostars and the regions where the filaments are unstable to gravitational collapse.
}
\label{aquila_polaris_filaments}
\end{figure*}

%||||||||||||||||||||||||||||||||||||||||||||||||||||||||||||||||||||||||||||||||||||||||||||||||||||||||||||||||||||||||||||||||||

\section{Main results and analysis}
\label{results}

The $Herschel$ images of the Aquila Rift and Polaris flare regions 
exhibit extensive filamentary structure, as well as numerous dense cores 
situated along these filaments (see Fig.~1, online Fig.~3,  
and Men'shchikov et al. 2010). 
A total of 541 starless cores ($\sim$~0.01--0.1~pc in size)
can be identified in the whole ($\ga 3.3^\circ \times 3.3^\circ $) Aquila field, most 
($> 60\% $) 
of which appear to be self-gravitating and prestellar in nature. 
The latter is inferred from: (1) a comparison of the core masses derived from the 
SPIRE/PACS spectral energy distributions with local values of the Jeans or Bonnor-Ebert (BE) mass also 
estimated from $Herschel$ data (see K\"onyves et al. 2010);   
(2)  the positions of the cores in a mass versus size diagram (cf. online Fig.~4), which are close 
to the loci expected for critical isothermal 
%Bonnor-Ebert 
BE spheres at gas temperatures $\sim$~7--20~K; 
and (3) 
%the fact that 
the mean column densities of the cores, which 
exceed the background column densities by a median factor $\sim 1.5$ as expected for critically 
self-gravitating BE 
%Bonnor-Ebert isothermal 
spheres. 
The shape of  the core mass function (CMF) derived for 
%the 391 prestellar cores of Aquila  
this sample of Aquila cores 
closely resembles the IMF (Fig.~2--left  -- see K\"onyves et al. 2010 for the CMF of the Aquila 
central region and 
Appendix~A for 
details on the derivation of core masses). 
In contrast, the 302 starless cores identified in the cirrus-like Polaris cloud with 
the same clump-finding algorithm (\textit{getsources} -- see Men'shchikov et al. 2010) appear to be mostly unbound (cf. online Fig.~4) and 
their mass distribution does {\it not} follow the IMF, with a peak at an order of magnitude smaller mass 
%$\sim 0.015\, M_\odot $ 
(Fig.~2--right). 
Only 5 of the Polaris cores are reasonably close to being gravitationally bound and  thus possibly prestellar in nature 
(see Ward-Thompson et al. 2010).
Between 45 and 60 Class~0 protostars are revealed by $Herschel$ in the Aquila field (Bontemps et al. 2010), 
while not a single protostar is detected in 
the Polaris region.

The Aquila filaments harboring embedded protostars and/or large concentrations of prestellar cores 
are all characterized by higher column densities ($N_{\rm H_2} \ga 10^{22}$~cm$^{-2}$), suggesting that 
they are  
gravitationally unstable (see Fig.~1--left).
In contrast, both the Polaris filaments and the quiescent, non-star-forming filaments observed in Aquila have much lower 
column densities (up to a few $10^{21}$~cm$^{-2}$), 
suggesting they are gravitationally stable (see Fig.~1--right and Sect.~4 below). 

\begin{figure*}
\centering
%\centerline{\resizebox{0.49\hsize}{!}{\includegraphics[angle=270]{14666fig2a.eps}}
%            \hspace*{0.2cm}\resizebox{0.49\hsize}{!}{\includegraphics[angle=270]{14666fig2b.eps}}}
\centerline{\resizebox{0.49\hsize}{!}{\includegraphics[angle=270]{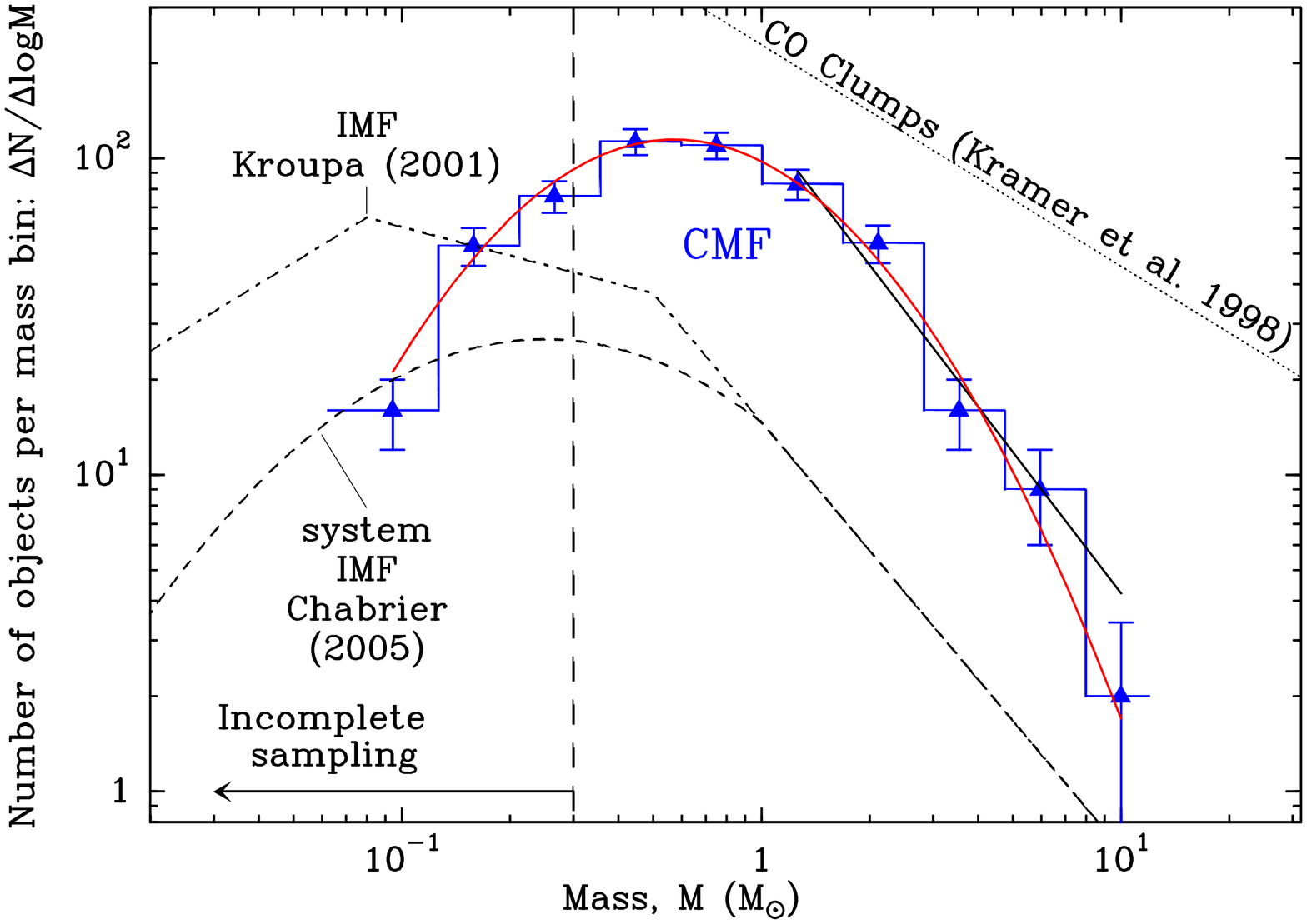}}
            \hspace*{0.2cm}\resizebox{0.49\hsize}{!}{\includegraphics[angle=270]{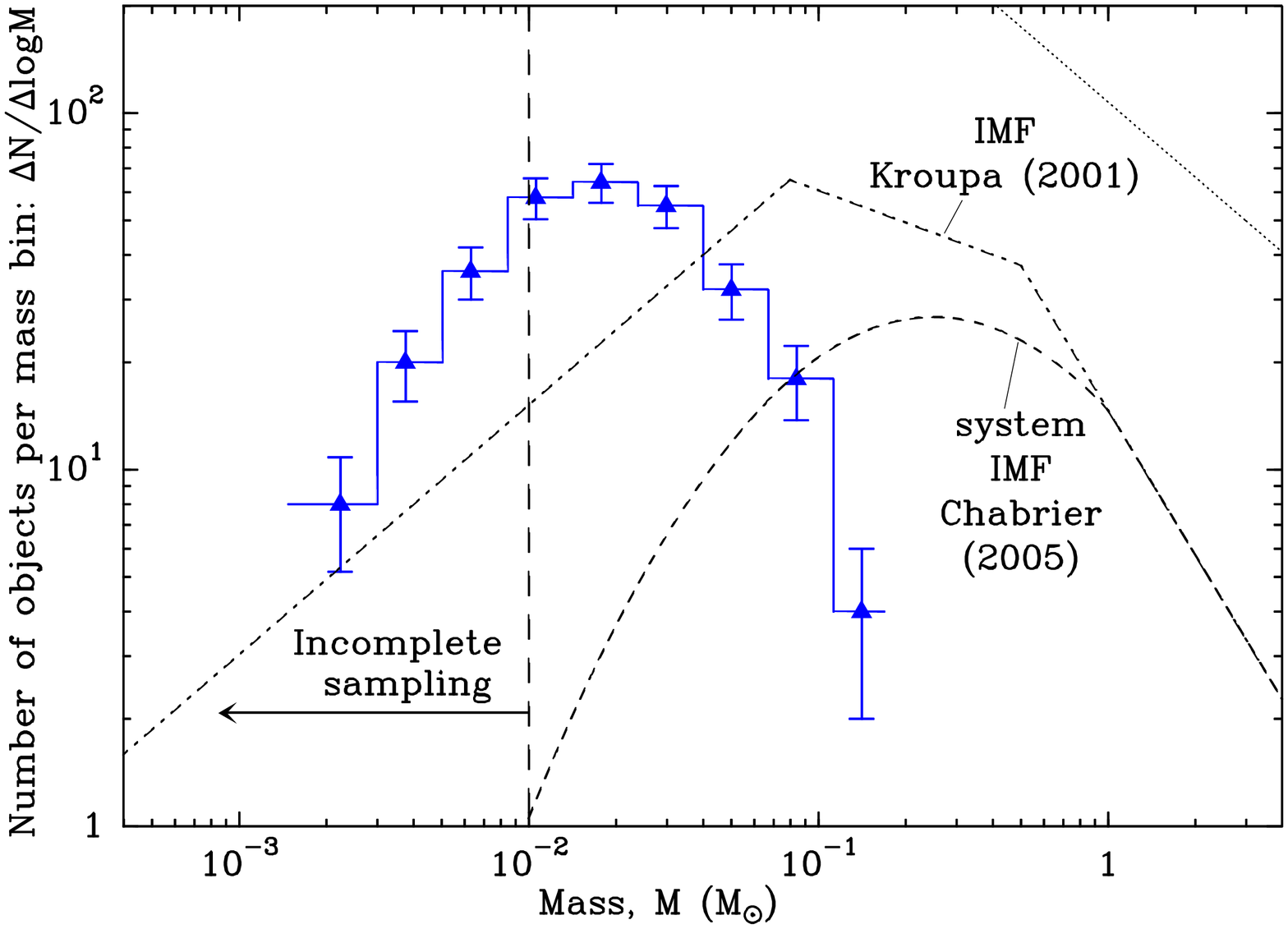}}}            
%\centerline{\resizebox{0.49\hsize}{!}{\includegraphics[angle=270]{../spire/sf_sag/sd_phase/spire_aquila/aquila_entire_cmf_nostruc.eps}}
%            \hspace*{0.2cm}\resizebox{0.49\hsize}{!}{\includegraphics[angle=270]{../spire/sf_sag/sd_phase/spire_polaris/polaris_cmf.eps}}}
\caption{Core mass functions (blue histograms with error bars) derived from our SPIRE/PACS observations 
of the Aquila (left) and Polaris (right) regions, which reveal of total of 541 candidate prestellar cores and 302 starless cores, respectively. 
A lognormal fit 
(red curve)
and a power-law fit 
(black solid line) 
to the high-mass end of the Aquila CMF are superimposed in the left panel. 
The power-law fit has a slope of $-1.5 \pm 0.2$ (compared to a Salpeter slope of $-1.35$ in this d$N$/dlog$M$ format), 
while the lognormal fit peaks at $\sim 0.6\, M_\odot $ and has a standard deviation of $\sim 0.43 $ in log$_{10}M$.
The IMF of single stars (corrected for binaries -- e.g., Kroupa 2001),  the IMF of multiple systems (e.g., Chabrier 2005), 
and the typical mass spectrum of CO clumps (e.g., Kramer et al. 1998) 
are also shown for comparison.
Note the remarkable similarity between the Aquila CMF and the stellar IMF, suggesting a $\sim$ one-to-one 
correspondence between core mass and star/system mass with $M_{\star \rm sys} = \epsilon M_{\rm core} $ 
and $\epsilon \approx $~0.4 in Aquila. 
}
\label{cmf_aquila_polaris}
\end{figure*}

\onlfig{3}
{
\begin{figure*} 
\centering
%\centerline{\resizebox{0.5\hsize}{!}{\includegraphics{14666fig3a.eps}}
%           \hspace*{0.1cm}\resizebox{0.47\hsize}{!}{\includegraphics{14666fig3b.eps}}}
\centerline{\resizebox{0.5\hsize}{!}{\includegraphics{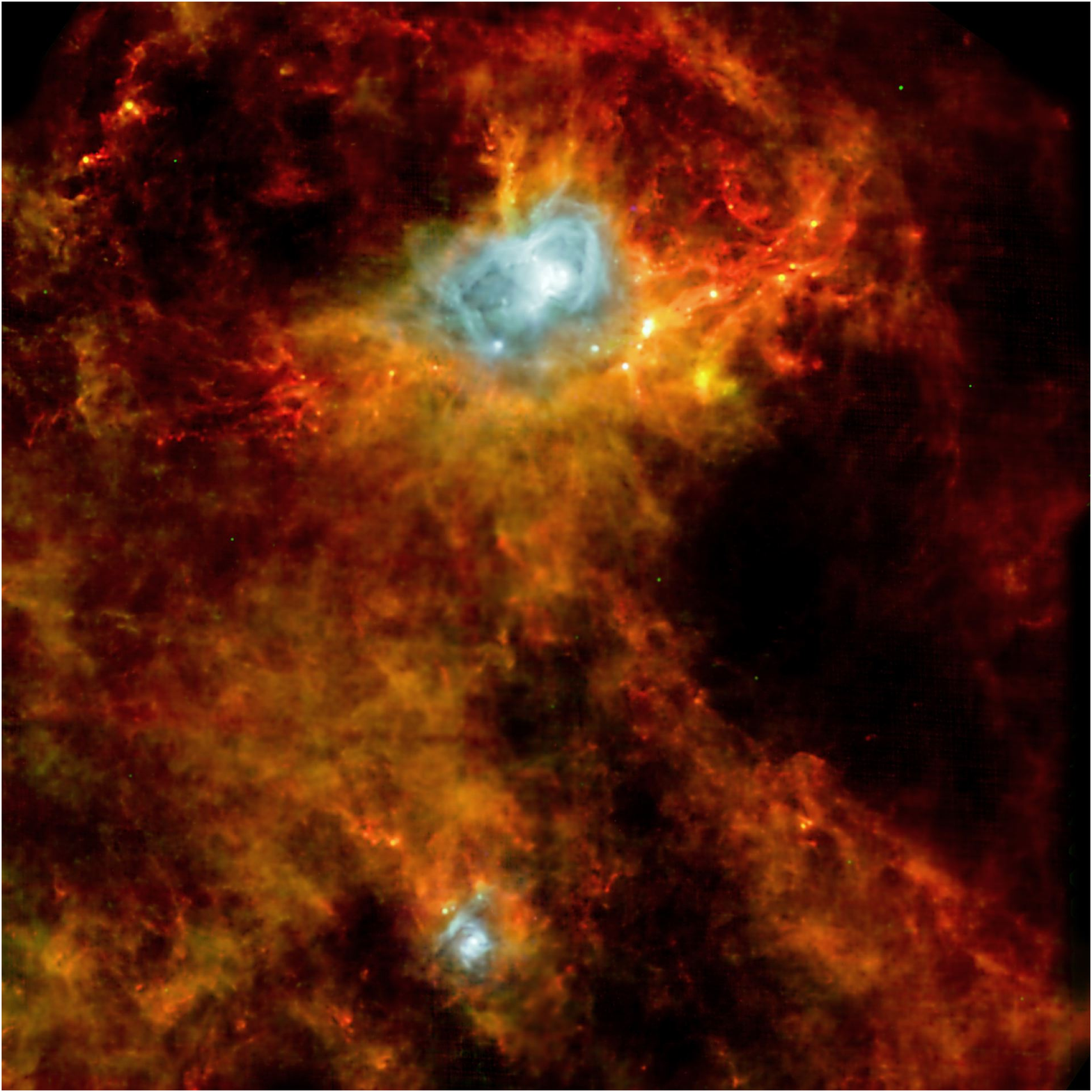}}
           \hspace*{0.1cm}\resizebox{0.47\hsize}{!}{\includegraphics{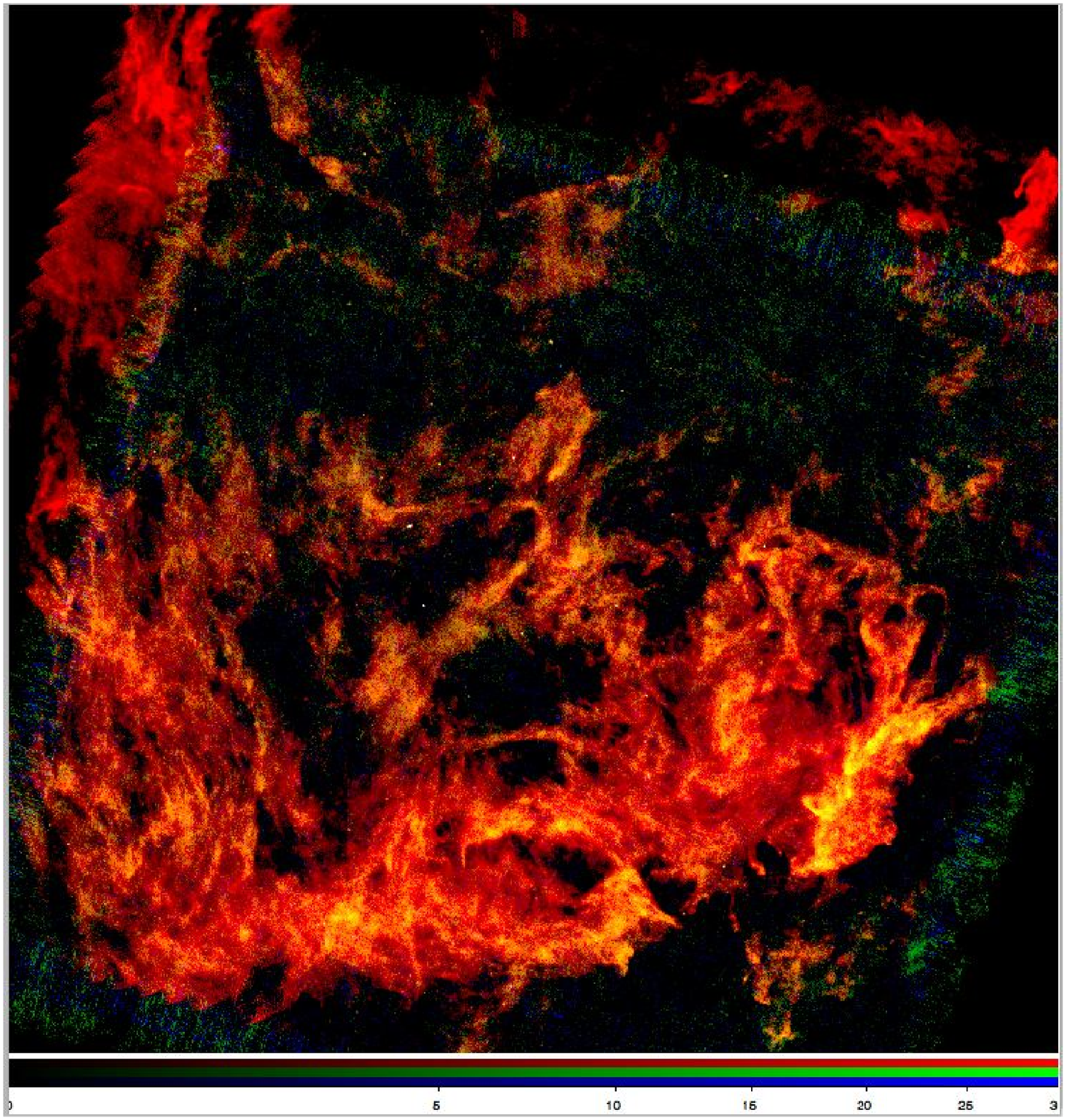}}}           
%\centerline{\resizebox{0.5\hsize}{!}{\includegraphics{../spire/sf_sag/sd_phase/PR_images/Aquila/HERSCHEL_Aquila_00_1930px_Bis_H_AandA.eps}}
%           \hspace*{0.1cm}\resizebox{0.47\hsize}{!}{\includegraphics{../spire/sf_sag/sd_phase/aa_filaments/polaris-spire-160-70.eps}}}
\caption{Composite 3-color images of the $\sim 11$~deg$^2$ Aquila field (\emph{left}) and the $\sim 8 $~deg$^2$ Polaris  field (\emph{right})
         produced from our PACS/SPIRE parallel-mode images at 70, 160, and 500\,{\um}. 
         The color coding of the Aquila image  is such that Red = SPIRE 500 $\mu $m,  Green = PACS 160 $\mu$m, Blue = PACS 70~$\mu$m.
         For the Polaris image, Red = combination of the three SPIRE bands, Green = PACS 160 $\mu$m, Blue = PACS 70~$\mu$m.
         The Aquila composite image was also the first release of ``OSHI'', ESA's Online Showcase of $Herschel$ Images 
         (cf. http://oshi.esa.int and http://www.esa.int/SPECIALS/Herschel/SEMT0T9K73G\_0.html).} 
\label{aquila.polaris.composites}
\end{figure*}
}

\onlfig{4}
{
\begin{figure*}
\vspace*{1.0 cm}
\centering
%\centerline{\resizebox{0.5\hsize}{!}{\includegraphics[angle=270]{14666fig4a.eps}}
%            \resizebox{0.5\hsize}{!}{\includegraphics[angle=270]{14666fig4b.eps}}}
\centerline{\resizebox{0.5\hsize}{!}{\includegraphics[angle=270]{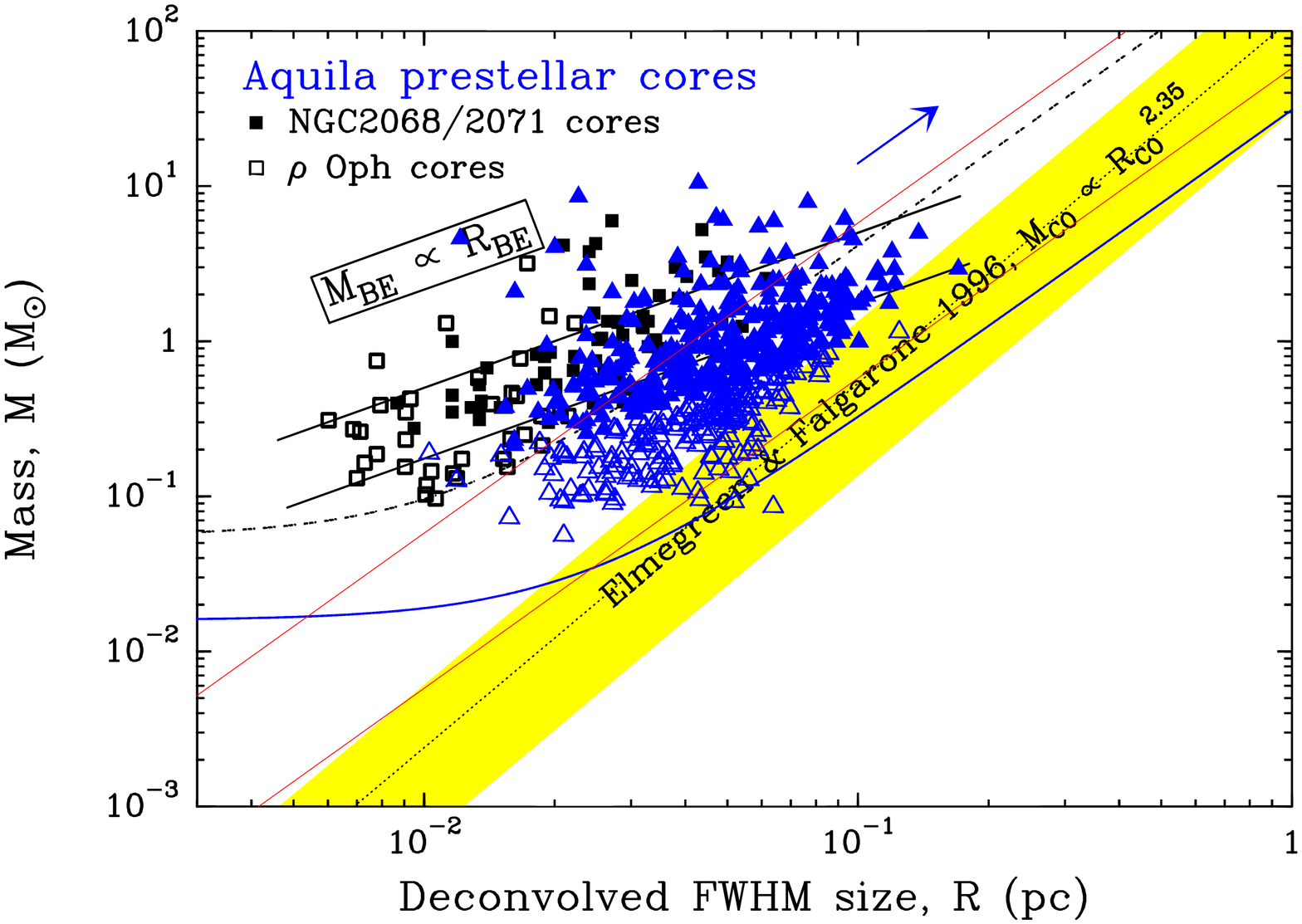}}
            \resizebox{0.5\hsize}{!}{\includegraphics[angle=270]{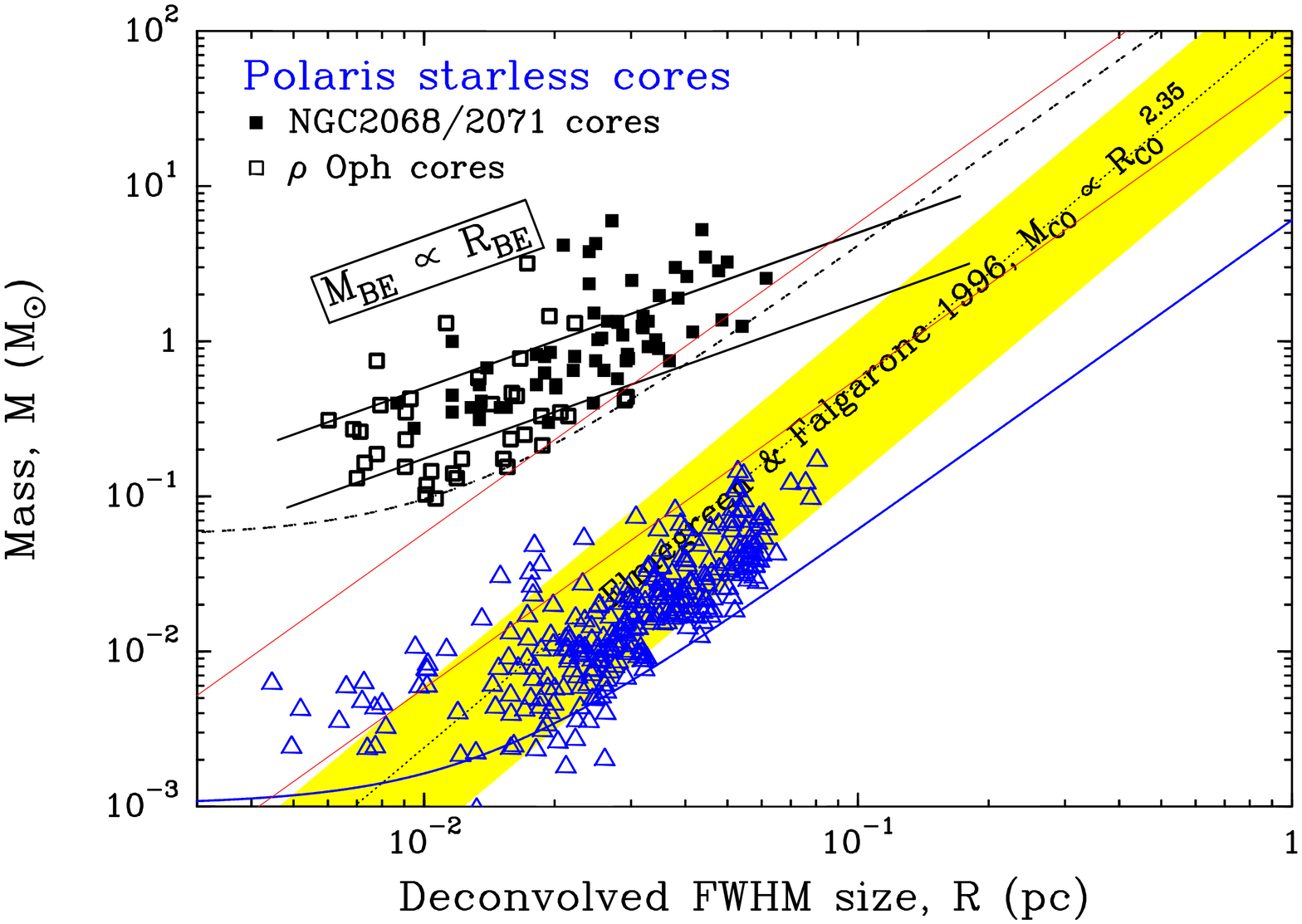}}}            
\caption{Mass vs. size diagrams for the starless cores detected with $Herschel$-SPIRE/PACS in Aquila (left) and Polaris (right) (blue triangles). 
The masses were derived as explained in Appendix~A (see also K\"onyves et al. 2010) and the sizes were measured at 250~$\mu$m. 
For reference, the locations of the (sub)mm continuum prestellar cores identified by Motte et al. (1998, 2001) in 
$\rho$~Oph and NGC2068/2071, respectively, are shown, along with the correlation observed 
for diffuse CO clumps (shaded band -- cf. Elmegreen \& Falgarone 1996).
%Note how the Aquila cores lie close to 
%cluster around 
The two black solid lines mark the loci of critically self-gravitating isothermal Bonnor-Ebert
spheres at $T = 7$~K and $T = 20$~K, respectively.
%, suggesting most of them are self-gravitating and prestellar in nature.
The 341 cores classified as prestellar by K\"onyves et al. (2010), out of a total of 541 starless cores in Aquila, 
are shown as filled triangles in the left panel.
%based on a comparison of their observed masses with estimates of the local Bonnor-Ebert mass.
The 302 starless cores of Polaris lie much below the two Bonnor-Ebert lines in the right panel, suggesting they all are unbound.
The red solid lines are two lines of constant mean column density at $N_{H_2} = 10^{21}$~cm$^{-2}$ and $N_{H_2} = 10^{22}$~cm$^{-2}$, 
respectively. The $5\sigma$ detection threshold at $d = 150$~pc of existing ground-based (sub)mm
(e.g., SCUBA) surveys as a function of size is shown by the dashed curve.
(The shape of this curve reflects a constant sensitivity to column density 
until source size approaches the beam size.)  
The $5\sigma$ detection thresholds of our SPIRE 250~$\mu$m observations, given the estimated levels 
of cirrus noise (cf. Sect.~2) and assumed distances of 260~pc for 
Aquila and 150~pc for Polaris, are shown by the blue curves.
The blue arrow in the left panel indicates how the positions of the $Herschel$ cores 
(and the blue curve marking the detection threshold)  
would move in the diagram 
if a distance of 400~pc were adopted for Aquila instead of 260~pc (see Appendix~A).
}
\label{mass_vs_size}
\end{figure*}
}

%||||||||||||||||||||||||||||||||||||||||||||||||||||||||||||||||||||||||||||||||||||||||||||||||||||||||||||||||||||||||||||||||||

%\newpage
%%%%%%%%%%%%%%%%%%
%%
\section{Discussion and conclusions}
\label{dis}

The $Herschel$ early results obtained toward the Aquila star-forming complex confirm that the prestellar CMF 
resembles the stellar IMF, using data with already a factor of $\sim $ 2 to 9 better counting statistics 
than earlier (sub)-millimeter continuum or near-IR extinction surveys 
(e.g., Motte et al. 1998, Johnstone et al. 2000, Enoch et al. 2006, Stanke et al. 2006, Nutter \& Ward-Thompson 2007, Alves et al. 2007). 
The close resemblance between the Aquila CMF and the IMF in both shape and mass scale suggests that, as a class, 
the self-gravitating prestellar cores identified in far-IR/submilimeter continuum imaging studies such as the present $Herschel$ survey
may form stars on a one-to-one basis, with a fixed and relatively high local efficiency, 
i.e., $\epsilon_{\rm core} \equiv M_\star /M_{\rm core} \sim 20-40\%$ in Aquila.
This is consistent with theoretical models according to which the stellar IMF is 
in large part determined by pre-collapse cloud fragmentation, prior to the protostellar accretion phase 
(cf. Larson 1985; Padoan \& Nordlund 2002; Hennebelle \& Chabrier 2008).  
There are several caveats to this simple picture (cf. discussion in Andr\'e et al. 2009),
and detailed analysis 
of the data from the whole Gould belt survey will be required to fully characterize the CMF--IMF relationship 
and, e.g., investigate possible variations in the efficiency $\epsilon_{\rm core}$ with environment.
It is nevertheless already clear that one of the keys to the problem of the origin of the IMF  
lies in a good understanding of the 
formation process of prestellar cores, 
even if additional processes, such as rotational subfragmentation of prestellar 
cores into binary/multiple systems (e.g., Bate et al. 2003), 
probably also play an important role.

Our $Herschel$ initial results also provide {\it key insight into the core formation issue}. 
They support an emerging picture (see also Myers 2009) 
according to which complex networks of long, thin filaments 
form first within molecular clouds, possibly as a result of interstellar MHD turbulence, and then the densest 
filaments fragment into a number of prestellar cores via gravitational instability. 
That the formation of filaments in the 
diffuse ISM represents the first step toward core/star formation is suggested by the filaments {\it already} being omnipresent 
in a diffuse, non-star-forming cloud such as Polaris (cf. Fig.~1--right, Men'shchikov et al. 2010, and 
Miville-Desch\^enes et al. 2010). The second step appears to be the gravitational fragmentation of a subset of the filaments 
into self-gravitating cores. This observationally-driven scenario can be placed on a stronger footing by comparing our  
$Herschel$ results to existing models of filamentary cloud fragmentation.
Inutsuka \& Miyama (1992, 1997) showed that an unmagnetized 
isothermal filament is unstable to axisymmetric perturbations if the 
line mass or mass per unit length, $M_{\rm line}$, of the filament is larger than the critical value required for equilibrium, 
$M_{\rm line, crit}^{\rm unmag} = 2\, c_s^2/G$,
where $c_s$ is the isothermal sound speed. 
Remarkably, the critical line mass 
only depends on gas temperature (Ostriker 1964) and is modified by only a factor of order unity for filaments 
with realistic levels of magnetization (Fiege \& Pudritz 2000). 
Figure~1 shows maps of the mass per unit length, expressed in approximate units of the 
critical line mass, for the filaments detected by $Herschel$ in Aquila and Polaris. 
These maps were constructed from the column density maps derived from our SPIRE/PACS images 
%(cf. K\"onyves et al. 2010) 
(see Appendix A) 
by multiplying the local column density of the filaments by their measured typical width 
(FWHM $\sim 14000$~AU in Aquila and $\sim 9000$~AU in Polaris, before
deconvolution -- cf. Men'shchikov et al. 2010). 
In Fig.~1, the critical line mass  
corresponding to a gas temperature of 10~K ($M_{\rm line, crit}^{\rm 10K, unmag} \approx  15\, M_\odot$/pc)
was adopted throughout the fields, but nearly 
identical results are obtained if the dust temperature maps derived from the $Herschel$ images (cf. Bontemps et al. 2010) are 
used instead. The results (cf. Fig.~1) show that most ($> 60\% $) of the bound prestellar cores  identified in Aquila are  
concentrated in  {\it supercritical} filaments with $M_{\rm line} > M_{\rm line, crit}^{\rm unmag} $. 
Furthermore, virtually all 
supercritical filaments harbor prestellar cores and/or embedded protostars, in agreement with the view that they are 
collapsing and actively forming stars at the present time. In particular, this is the case for the Aquila main filament, which has 
$M_{\rm line} > 5 \times M_{\rm line, crit}^{\rm unmag} $ and is associated with a very rich protocluster (Serpens South -- see Gutermuth et al. 2008, Bontemps et al. 2010). 
In contrast, the {\it subcritical} filaments with $M_{\rm line} < M_{\rm line, crit}^{\rm unmag} $ are 
generally 
devoid of prestellar cores and protostars, which is consistent 
with the view that they are gravitationally stable, hence neither collapsing nor forming stars. 
{\it All} of the Polaris filaments appear to be in the subcritical regime, the maximum observed value of 
the stability parameter being $M_{\rm line}/M_{\rm line, crit}^{\rm unmag}  \sim 0.45 $, and it is unclear whether they will evolve into the 
unstable regime or not.

It is noteworthy that the critical line mass approximately corresponds to a critical column density $N_{\rm H_2, crit} \sim 10^{22}$~cm$^{-2}$,  
i.e., to a critical visual extinction $A_{\rm V, crit} \sim 10$. 
Our $Herschel$ findings thus provide an {\it explanation} of the 
visual extinction threshold for core formation found by 
earlier ground-based studies (e.g., Onishi et al. 1998, Johnstone et al. 2004). Prestellar cores are only observed above a threshold $A_{\rm V, crit} $ 
because they form out of a filamentary background and only the supercritical, gravitationally unstable filaments are able 
to fragment into bound cores. 

Confirming and refining this scenario for core formation will require the results of the entire $Herschel$ survey, 
as well as follow-up (sub-)millimeter dust polarimetry and molecular line observations to, e.g., 
%clarify the role of magnetic fields and turbulence in 
clarify the roles of magnetic fields, turbulence, and gravity in 
forming the filaments.   
Our initial results are nevertheless extremely encouraging. 
Extrapolating from the $\ga 500$ prestellar cores and $\ga 45$ Class~0 protostars identified in 
the $\sim 19$~deg$^2$ covered by the Aquila and Polaris fields, 
we expect that the $\sim 160$~deg$^2$ Gould belt survey will reveal a total of about 4500 prestellar cores, 
including a large number of candidate pre-brown dwarfs in the nearest ($d \sim 150$~pc) clouds, and more than 350 Class~0 protostars. 
This will provide a unique database, including the southern hemisphere, for follow-up 
high-resolution molecular line/dust continuum studies of the physics of individual cores and protostars with ALMA.

%||||||||||||||||||||||||||||||||||||||||||||||||||||||||||||||||||||||||||||||||||||||||||||||||||||||||||||||||||||||||||||||||||

\begin{acknowledgements}

SPIRE has been developed by a consortium of institutes led by
Cardiff Univ. (UK) and including Univ. Lethbridge (Canada);
NAOC (China); CEA, LAM (France); IFSI, Univ. Padua (Italy);
IAC (Spain); Stockholm Observatory (Sweden); Imperial College
London, RAL, UCL-MSSL, UKATC, Univ. Sussex (UK); Caltech, JPL,
NHSC, Univ. Colorado (USA). This development has been supported
by national funding agencies: CSA (Canada); NAOC (China); CEA,
CNES, CNRS (France); ASI (Italy); MCINN (Spain); SNSB (Sweden);
STFC (UK); and NASA (USA).
PACS has been developed by a consortium of institutes led by MPE (Germany) and including UVIE 
(Austria); KU Leuven, CSL, IMEC (Belgium); CEA, LAM (France); MPIA (Germany); 
INAF-IFSI/OAA/OAP/OAT, LENS, SISSA (Italy); IAC (Spain). This development has been supported by the 
funding agencies BMVIT (Austria), ESA-PRODEX (Belgium), CEA/CNES (France), DLR (Germany), 
ASI/INAF (Italy), and CICYT/MCYT (Spain).

\end{acknowledgements}

%||||||||||||||||||||||||||||||||||||||||||||||||||||||||||||||||||||||||||||||||||||||||||||||||||||||||||||||||||||||||||||||||||

%||||||||||||||||||||||||||||||||||||||||||||||||||||||||||||||||||||||||||||||||||||||||||||||||||||||||||||||||||||||||||||||||||

\Online

\begin{appendix} %First online appendix
\section{Derivation of core/filament properties and effects of distance uncertainties}

As described in more detail in a companion paper by K\"onyves et al. (2010) on Aquila, the 
masses of the cores identified in the $Herschel$ images with the \textit{getsources} algorithm (see Men'shchikov et al. 2010) 
were derived by fitting grey-body functions to the spectral energy distributions (SEDs) constructed from the integrated 
flux densities measured with SPIRE/PACS for each core. We assumed the dust opacity law 
$\kappa_{\nu} = 0.1~(\nu/1000~{\rm GHz})^2$~cm$^2$/g, where $\nu $ denotes frequency and 
$\kappa_{\nu}$ is the dust opacity per unit (gas~$+$~dust) mass column density.
This dust opacity law, which is very similar to that advocated by Hildebrand (1983), is consistent 
with the value $\kappa_{\rm 1.3mm} = 0.005$~cm$^2$/g adopted for starless cores in numerous 
earlier studies (e.g., Motte et al. 1998).
Ignoring any systematic distance effect (see below), the core mass uncertainties are dominated by the uncertainty in $\kappa_\nu$,  
typically a factor of $\sim 2$. 
Cores were classified as either protostellar or starless based on the presence or absence of significant PACS emission 
at 70~$\mu$m, respectively (cf. Bontemps et al. 2010 and Dunham et al. 2008). 
In the Polaris field, the cirrus noise level is so low (cf. Fig.~4--right) that we cannot exclude 
that a fraction of the 302 candidate starless cores extracted with \textit{getsources} correspond 
to background galaxies.

A column density map was derived for each region from the $Herschel$ images smoothed to the SPIRE  500~$\mu$m 
resolution (36.9\arcsec ~FWHM) using a similar SED fitting procedure on a pixel by pixel basis 
(see, e.g., Figs.~1 and~6 of K\"onyves et al. 2010 for Aquila). 
To obtain the maps of the filamentary background shown in Fig.~1 of this paper, we then performed 
a ``morphological component analysis'' decomposition (e.g., Starck et al. 2003) of the original column density maps
on curvelets and wavelets. The curvelet component images shown in Fig.~1 provide a good measurement of the column 
density distribution of the filamentary background after subtraction of the compact sources/cores since the latter are contained 
in the wavelet component. 
We estimate that these column density maps are accurate to within a factor of $\sim 2$. 
The scaling in terms of the mass per unit length along the filaments is more uncertain, however, as it depends on distance 
(see below) and would in principle require a detailed analysis of the radial profiles of the filaments, which is beyond the 
scope of the present letter. Here, we simply assumed that the filaments had a Gaussian radial column density profile 
and multiplied the surface density maps by $\sqrt{\frac{2\pi}{8\, {\rm ln} 2}} \times W \approx 1.06\, W $, where $W$ 
is the typical FWHM width of the filaments. We assumed a mean molecular weight of $\mu = 2.33$. 
At this stage, the correspondence between the critical line mass of the filaments, $M_{\rm line, crit}^{\rm unmag}$, and the visual 
extinction threshold, $A_{\rm V, crit} \sim 10$ (see Sect.~4), is thus accurate to at best a factor of $\ga 2$. 

There is some ambiguity concerning the distance to the Aquila Rift region. 
A number of arguments, presented in a companion paper by Bontemps et al. (2010), 
suggest that the whole region corresponds to a coherent cloud complex at $d_{-} = 260$~pc 
(see also Gutermuth et al. 2008), which is the default distance adopted in the present paper for Aquila. 
However, other studies in the literature (see references in Bontemps et al. 2010) 
place the complex at a larger distance, $d_{+} = 400$~pc. 
It is thus worth discussing briefly how our Aquila results would be affected 
if we adopted the larger distance estimate,  $d_{+} $, instead of $d_{-}$. 
The core mass estimates, which scale as $S_{\nu}\, d^2/[B_{\nu}(T_{\rm d})\, \kappa_{\nu}] $ where 
$S_{\nu}$ is integrated flux density and $B_{\nu}(T_{\rm d})$ is the Planck function,  
would systematically increase by a factor of  2.4. 
This would shift the CMF shown in Fig.~2--left to the right and 
thus lower the efficiency $\epsilon_{\rm core}$ from $\sim $ 20--40\% to $\sim$ 10--20\%.
In the mass versus size diagram of Fig.~4, the cores would move upward as indicated in the left panel
of the figure, which would {\it increase} the fraction of candidate bound cores in Aquila from $63\% $ to $81\%$. 
%(see also K\"onyves et al. 2010). 
The column density map of the Aquila filaments shown in Fig.~1 would be unchanged, but 
the scaling in terms of the mass per unit length  along the filaments would change by $\sim 50\% $ 
upward, since the physical width of the filaments would increase by $\sim 50\% $.
%from $\sim 9000$~AU to $\sim 14000$~AU. 
In other words, the highlighted regions in Fig.~1--left, where the mass per unit length  of the filaments 
exceeds half the critical value, would slightly {\it expand}, increasing the contrast with the Polaris 
filaments and improving the correspondence between the spatial distribution of the 
prestellar cores/protostars in Aquila and that of the gravitationally unstable filaments.
To summarize, our main conclusions do not depend strongly on the adopted distance.
% and, if anything, 
%would be strengthened if a distance of 400~pc were adopted for the Aquila region.

\end{appendix}

%%\begin{appendix}
%%
%%\section{Online figures}
%%
%%\end{appendix}


\normalsize
\begin{thebibliography}{}

\bibitem[Alves et al.(2007)]{Alves07} Alves, J. F., Lombardi, M., \& Lada, C. J. 2007, \aap, 462, L17
\bibitem[Andr{\' e} et al.(2009)]{Andre09} Andr\'e, P., Basu, S., \& Inutsuka, S.-I. 2009, in Structure Formation in Astrophysics, Ed. G. Chabrier, 
Cambridge University Press, p. 254
\bibitem[Andr{\' e} \& Saraceno(2005)]{Andre05} Andr\'e, P. , \& Saraceno, P. 2005, in The Dusty and Molecular Universe: A Prelude to $Herschel$ and ALMA, ESA SP-577,  p. 179
\bibitem[Bate et al.(2003)]{Bate03} Bate, M. R., Bonnell, I. A., \& Bromm, V. 2003, \mnras, 339, 577
\bibitem[Bontemps et al.(2010)]{Bontemps10} Bontemps, S., Andr\'e, Ph., K\"onyves, V. et al. 2010, \aap, this volume
\bibitem[Chabrier(2005)]{chabrier05} Chabrier, G. 2005, in The Initial Mass Function 50 years later, Eds. E. Corbelli et al., p.41
\bibitem[Dunham et al.(2008)]{Dunham08} Dunham, M.M., Crapsi, A., Evans, N.J. et al.  2008, \apjs, 179, 249
\bibitem[Elmegreen \& Falgarone(1996)]{ElmFalg96} Elmegreen, B.G., \& Falgarone, E. 1996, \apj, 471, 816
\bibitem[Enoch et al.(2006)]{Enoch06} Enoch, M. L., Young, K. E., Glenn, J., Evans, N. J. et al.  2006, \apj, 638, 293
\bibitem[Fiege \& Pudritz(2000)]{Fiege00} Fiege, J.D., \& Pudritz, R.E. 2000,  \mnras, 311, 85
\bibitem[Griffin et al.(2010)]{Griffin10} Griffin, M., Abergel, A., Abreu, A. et al. 2010, \aap, this volume
\bibitem[Guillout(2001)]{Guillout01} Guillout, P. 2001, in From Darkness to Light, Eds. T. Montmerle \& P. Andr\'e,
ASP Conf. Ser., 243, p. 677
\bibitem[Gutermuth et al.(2008)]{Gutermuth08} Gutermuth, R.A., Bourke, T.L., Allen, L.E. et al. 2008, \apj, 673, L151
\bibitem[Heithausen et al.(2010)]{Heithausen02} Heithausen, A. et al. 2002, \aap, 383, 591
\bibitem[Hennebelle \& Chabrier(2008)]{Hennebelle08} Hennebelle, P., \& Chabrier, G. 2008, \apj, 684, 395
\bibitem[Hidebrand(1983)]{Hildebrand83} Hildebrand, R.H. 1983, \qjras, 24, 267
\bibitem[Inutsuka \& Miyama(1992)]{Inutsuka92} Inutsuka, S-I, \& Miyama, S.M. 1992, \apj, 388, 392
\bibitem[Inutsuka \& Miyama(1997)]{Inutsuka97} Inutsuka, S-I, \& Miyama, S.M. 1997, \apj, 480, 681
\bibitem[Johnstone et al.(2000)]{Johnstone00} Johnstone, D., Wilson, C.D., Moriarty-Schieven, G. et al. 2000, \apj, 545, 327
\bibitem[Johnstone et al.(2004)]{Johnstone04} Johnstone, D., Di Francesco, J., \& Kirk, H. 2004, \apj, 611, L45
\bibitem[K\"onyves et al.(2010)]{Konyves10} K\"onyves, V., Andr\'e, Ph., Men'shchikov, A. et al. 2010, \aap, this volume
\bibitem[Kramer et al.(1998)]{Kramer98} Kramer, C., Stutzki, J., Rohrig, R., Corneliussen, U. 1998, \aap, 329, 249
\bibitem[Kroupa(2001)]{kroupa01} Kroupa, P. 2001, \mnras, 322, 231
\bibitem[Larson(1985)]{Larson85} Larson, R. 1985, \mnras, 214, 379 
\bibitem[Men'shchikov et al.(2010)]{Menshch10} Men'shchikov, A., Andr\'e, Ph., Didelon, P. et al. 2010, \aap, this volume
\bibitem[Miville-Desch\^enes et al.(2010)]{Miville10} Miville-Desch\^enes, M.-A., Martin, P.G., Abergel, A. et al. 2010, \aap, this volume
\bibitem[Motte et al.(1998)]{Man98} Motte, F., Andr\'e, P., \& Neri, R. 1998, \aap, 365, 440
\bibitem[Motte et al.(2001)]{Motte01} Motte, F., Andr\'e, P., Ward-Thompson, D., \& Bontemps, S. 2001, \aap, 372, L41
\bibitem[Myers(2009)]{kroupa09} Myers, P.C.  2009, \apj, 700, 1609
%\bibitem[Nagai et al.(1998)]{nagai98} Nagai, T., Inutsuka, S.-I., \& Miyama, S. M. 1998, \apj, 506, 306
\bibitem[Nutter \& WardThompson(2007)]{nutter07} Nutter, D., \& Ward-Thompson, D. 2007, \mnras, 374, 1413
\bibitem[Onishi et al.(1998)]{onishi98} Onishi, T., Mizuno, A., Kawamura, A. et al. 
%, Ogawa, H., Fukui, Y. 
1998, \apj, 502, 296
\bibitem[{{Ostriker}(1964)}]{Ostriker1964} {Ostriker}, J. 1964, \apj, 140, 1056
\bibitem[Padoan \& Nordlund(2002)]{padoan02} Padoan, P. \& Nordlund, A. 2002, ApJ, 576, 870
\bibitem[Pilbratt et al.(2010)]{Pilbratt10} Pilbratt, G.L., Riedinger, J.R., Passvogel, T. et al. 2010, \aap, this volume
\bibitem[Poglitsch et al.(2010)]{Poglitsch10} Poglitsch, A., Waelkens, C., Geis, N. et al. 2010, \aap, this volume
\bibitem[Roy et al.(2010)]{Roy10} Roy, A., Ade, P.A.R., Bock, J.J. et al. 2010, \apj, 708, 1611
\bibitem[Stanke et al.(2006)]{Stanke06} Stanke, T., Smith, M. D., Gredel, R., \& Khanzadyan, T.\ 2006, \aap, 447, 609
\bibitem[Starck et al.(2003)]{Starck03} Starck, J. L., Donoho, D. L., Cand\`es, E. J.\ 2003, \aap, 398, 785
\bibitem[Ward-Thompson et al.(2010)]{Ward10} Ward-Thompson, D., Kirk, J.M., Andr\'e, P. et al. 2010, \aap, this volume


\end{thebibliography}
\end{document}